\documentclass[12pt]{iopart}
\usepackage{graphicx}% Include figure files
\usepackage{dcolumn}% Align table columns on decimal point
\usepackage{bm}% bold math
\usepackage{times}
\usepackage[colorlinks,citecolor=blue,linkcolor=red]{hyperref}
\usepackage{color}
\usepackage{ulem}
\begin{document}

\title[Loss revives bistable state near the exceptional point in a non-Hermitian photonic meta-molecule]{Loss revives bistable state near the exceptional point in a non-Hermitian microwave photonic meta-molecule}

%\author{Kejia Zhu, Yong Sun, Jie Ren \& Hong Chen}

%\address{School of Physics Science and Engineering, Tongji University, Shanghai 200092, China}
%\ead{yongsun@tongji.edu.cn \& xonics@tongji.edu.cn}

\author{Kejia Zhu}
%\email{1510042\_zhu@tongji.edu.cn}
 \address{School of Physics Science and Engineering, Tongji University, Shanghai 200092, China}%Lines break automatically or can be forced with \\
\author{Yong Sun}%
\ead{yongsun@tongji.edu.cn}
\address{School of Physics Science and Engineering, Tongji University, Shanghai 200092, China}%

\author{Jie Ren}
\ead{xonics@tongji.edu.cn}
\address{School of Physics Science and Engineering, Tongji University, Shanghai 200092, China}
\address{Center for Phononics and Thermal Energy Science, Tongji University, Shanghai 200092, China}
\address{China-EU Joint Center for Nanophononics, Tongji University, Shanghai 200092, China}
\address{Shanghai Key Laboratory of Special Artificial Microstructure Materials and Technology, Tongji University, Shanghai 200092, China}

\author{Hong Chen}%
%\email{hongchen@tongji.eud.cn}
\address{School of Physics Science and Engineering, Tongji University, Shanghai 200092, China}%
\address{Center for Phononics and Thermal Energy Science, Tongji University, Shanghai 200092, China}

\vspace{10pt}
\begin{indented}
\item[]January 2017
\end{indented}

\begin{abstract}
By exploring the extraordinary property of exceptional points (EPs) in non-Hermitian systems, we here demonstrate that losses can play constructive roles in controlling bistable states. We experimentally realize the EP in a non-Hermitian meta-molecule of coupled resonators in microwave regime. By increasing the loss, we first observe the bistable state suppression at the weak-dissipative regime, but then the bistable state recovery in strong-dissipative regime. Both the experimental and theoretical analysis demonstrate that the revival of bistable states results from the revival of the field intensity after the system encounters EPs, in spite of the increasing loss. Our results provide an alternative way to controlling and manifesting bistable systems, so as to achieve flexible photonic devices, not limited to the microwave regime.
\end{abstract}

% Uncomment for PACS numbers
%\pacs{00.00, 20.00, 42.10}
%
% Uncomment for keywords
%\vspace{2pc}
%\noindent{\it Keywords}: XXXXXX, YYYYYYYY, ZZZZZZZZZ
%
% Uncomment for Submitted to journal title message
%\submitto{\JPA}
%
% Uncomment if a separate title page is required
%\maketitle
% 
% For two-column output uncomment the next line and choose [10pt] rather than [12pt] in the \documentclass declaration
%\ioptwocol
%
\maketitle

\section{Introduction}

Naturally, dissipation is always eliminated or negeleted in normal physical systems due to its disadvantages such as wasting energy, reducing field intensity and increasing threshold. Recent years, however, researchers discovered nontrivial physics in non-Hermitian dissipative systems described by complex eigenvalues and nonorthogonal eigenstates that departure from a conventional Hermitian model~\cite{bender2007making}. When steering parameters of such a dissipative system, it encounters a different type of singularities where the eigenvalues and the corresponding eigenstates may coalesce by a square root branch point, which has been dubbed an exceptional point (EP) by Kato~\cite{kato1976perturbation,heiss2012physics,dembowski2004encircling}. Also, an EP may emerge in purely dielectric crystals\cite{yannopapas2013spontaneous}, even in topological photonic systems such as lattices of gyrotropic materials\cite{yannopapas2012non} or lattices of resonators coupled with metamaterial elements\cite{yannopapas2012topological}. Spectral coalesces crucially determine unusual transport properties~\cite{guo2009observation} and topological structure of eigenmodes~\cite{dembowski2001experimental,gao2015observation} in the vicinity of EPs such as counter-intuitive features of lasing~\cite{peng2014loss,brandstetter2014reversing,Chitsazi2014Experimental,miri2012large,feng2014single,hodaei2014parity,gentry2014dark}, chiral modes~\cite{dembowski2003observation} and repulsion of energy levels~\cite{heiss2000repulsion}. In particular, physics of EPs even have been associated with parity-time ($\mathcal{PT}$) symmetric~\cite{bender1998real}, where losses and gains are balanced~\cite{el2007theory,peng2014loss,ruter2010observation,klaiman2008visualization}, with effects such as coherent perfect absorption~\cite{sun2014experimental,fan2014tunable}, loss-induced transparency~\cite{guo2009observation} and unidirectional invisibility~\cite{regensburger2012parity}. Additionally, non-Hermitian stochastic dynamics have so far been studied in the context of microwave~\cite{dembowski2001experimental,dembowski2003observation,
dembowski2004encircling,bliokh2008coupling,
gao2015observation,cao2015dielectric,cerjan2016eigenvalue}, optical~\cite{gao2015observation,peng2014loss,sun2014experimental}, atomic~\cite{choi2010quasieigenstate,milner2001optical,kaplan2001observation} and electron waves~\cite{gao2015observation,akis1997wave,ponomarenko2008chaotic}.

Very recently, two coupled waveguides~\cite{ruter2010observation,klaiman2008visualization,brandstetter2014reversing,el2007theory} and resonators model~\cite{peng2014loss,sun2014experimental,Verslegers2012From} have attracted enormous attention as they can demonstrate properties of optical systems visually. Other recent studies in lasing regime have also given examples, like a pump-induced lasing death~\cite{liertzer2012pump}, spontaneous $\mathcal{PT}$ symmetry breaking~\cite{ruter2010observation} and enhancement of the laser linewidth~\cite{wenzel1996mechanisms}. Notably, the photonic ``meta-molecule'' in microwave regime constructed by metamaterial units (meta-atom) ~\cite{sun2014experimental,tan2014manipulating,Sun2015Dephasing} triggers a unique view in investigating the internal physics of such systems. 

In this Letter, we first show how to simulate an EP in a non-Hermitian microwave photonic meta-molecule made of coupled resonators. Then, by increasing the dissipative loss to a resonator solely, the system undergoes a non-ideal $\mathcal{PT}$ phase transition, which is shown clearly from the experimental transmission. With introducing the nonlinearity on one resonator, we observe the bistable state of this system, manifesting as the transmission hysteresis when scanning in power and frequency respectively. Further, evolution of measured intensity uncovers that the revival of the bistable state results from the revival of the field strength in this system, which also matches the theoretical and simulated results. The investigations enrich the manifestation of EPs physics, and help to realize such anomalous phenomena and offer a supporting platform for researching bistable state systems.

%\section{theoretical model and experiment}
\begin{figure}[b]
\includegraphics[width=0.85\linewidth]{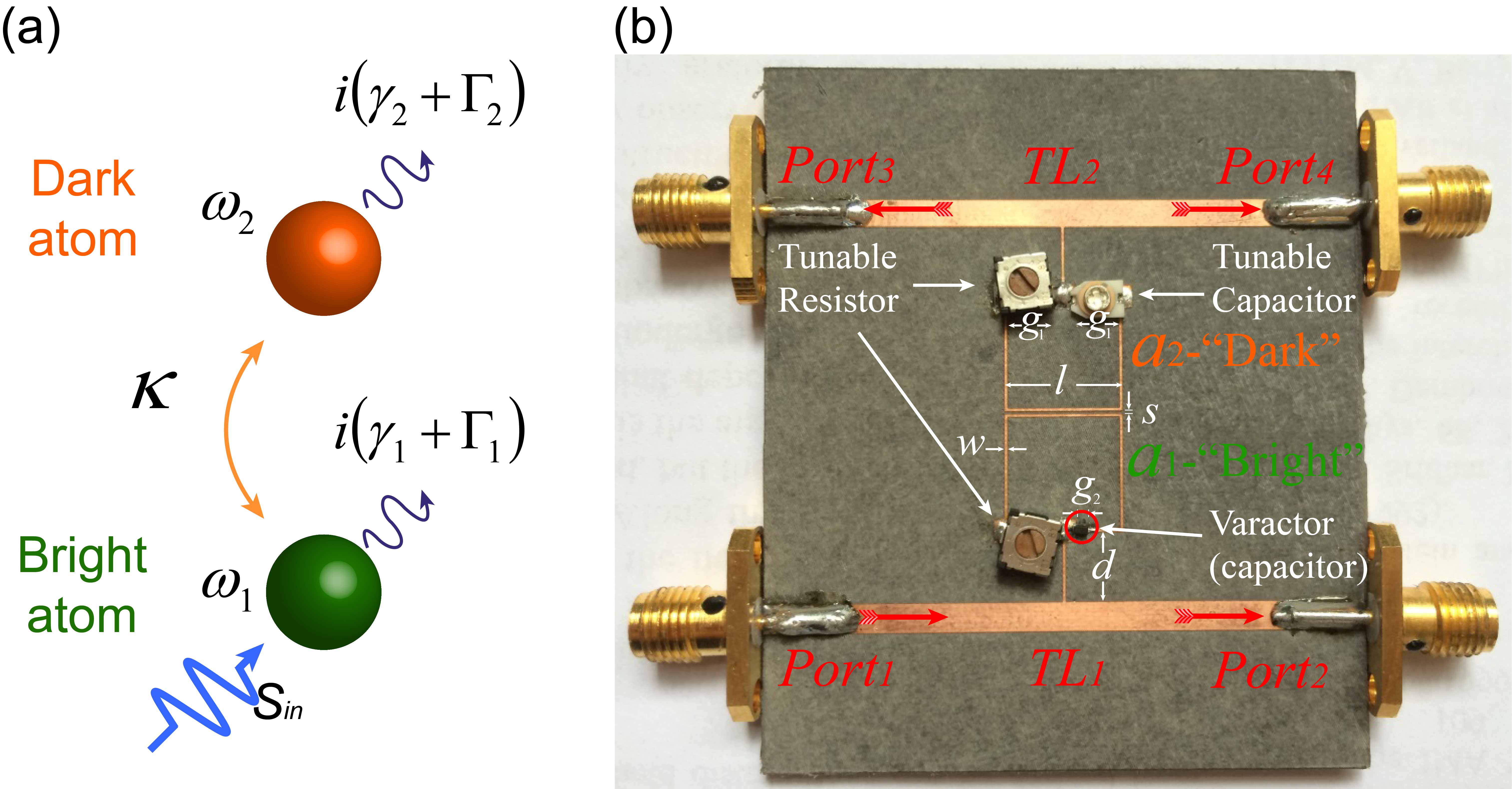}
\caption{\label{fig:sample} (a) Model of meta-molecule constructed by ``Bright" and ``Dark" meta-atoms. (b) Experimental sample constructed by coupled resonators $a_1$ (``Bright" meta-atom) and $a_2$ (``Dark" meta-atom) with the transmission line couplers $TL_1$ and $TL_2$.}
\end{figure} 

\section{Theoretical and experimental results}
\begin{figure}
\includegraphics[width=1.0\linewidth]{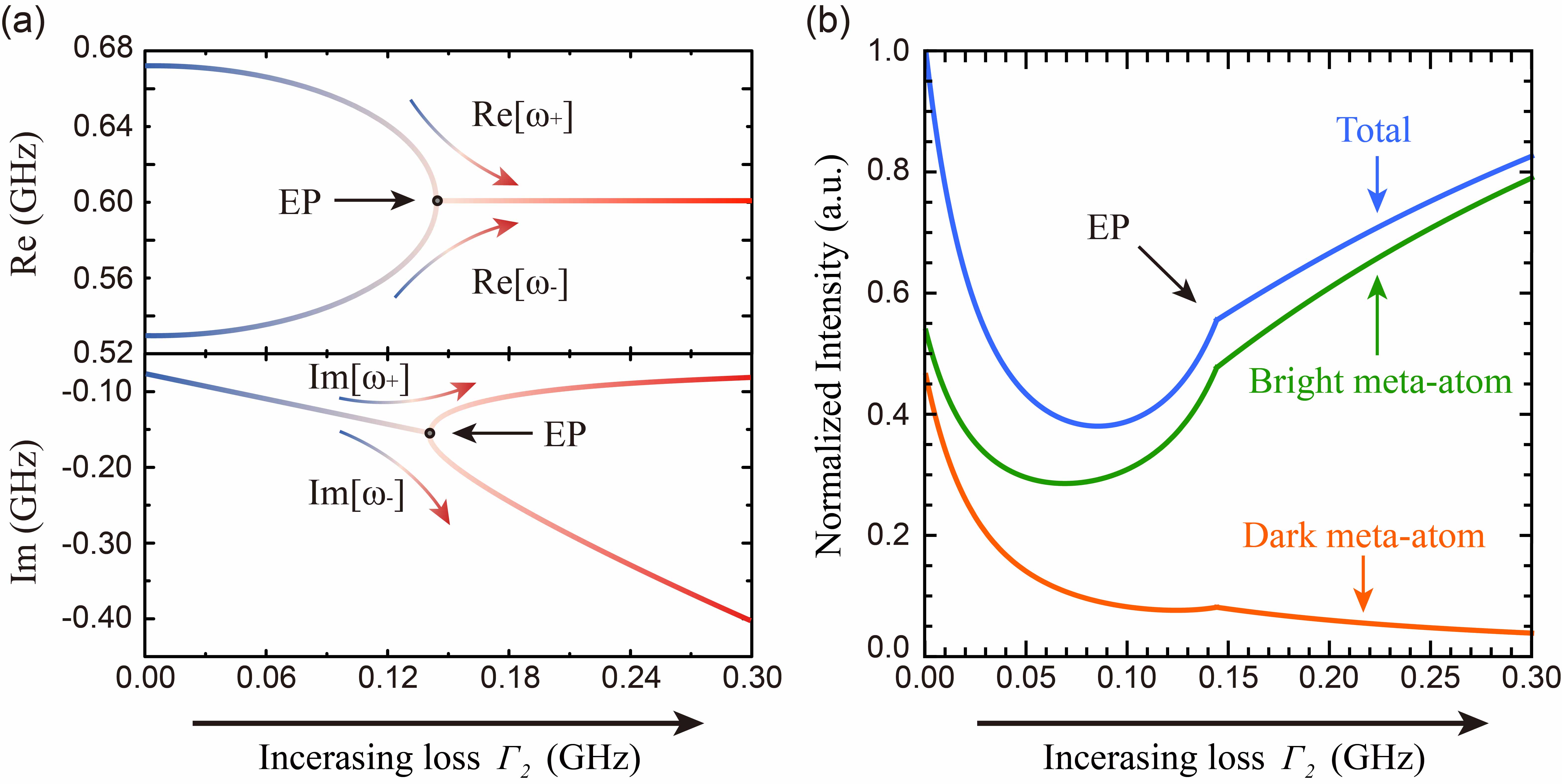}
\caption{\label{fig:calculated re&im intensity} (Color online) (a) Calculated evolution of real part and imaginary part of eigenfrequencies ($\omega_+$ and $\omega_-$) with increasing dissipative loss in the ``dark" meta-atom $\Gamma_2$. (b) Corresponding evolution of normalized field intensity in ``bright" meta-atom (green line), ``dark" meta-atom (orange line) and total meta-molecule (blue line). System parameters in the experiment are fit from measured data: the radiative losses of two meta-atoms $\gamma_1=0.032$ GHz, $\gamma_2=0.028$ GHz and the coupling strength $\kappa=0.07$ GHz. We also set $\Gamma_1=0$ GHz by tuning the resistor 1 to zero, $R_1=0$. The following studies use the same parameters if without specified.} 
\label{Figure2}
\end{figure}

Our system is based on a microwave photonic molecule, as shown in Fig.~\ref{fig:sample}(a). This molecule model is constructed by a pair of coupled atoms, ``bright" atom (the lower green sphere) and ``dark" atom (the upper orange sphere). The ``bright" atom with resonant frequency $\omega_1$ can be easily excited by incoming wave $S_{in}e^{-i{\omega}t}$, while the ``dark" atom with resonant frequency $\omega_2$ can only be excited by the near-field from the bright atom through a coupling parameter $\kappa$. When considering the radiative loss $\gamma_{1,2}$ and dissipative loss $\Gamma_{1,2}$ of the two atoms, meta-molecule can be regarded as a non-Hermitian system. To mimic the above-mentioned microwave photonic molecule exactly, we designed the experimental system with two coupled ring resonators in Fig.~\ref{fig:sample}(b). %There are two coupled ring resonators, $a_1$ and $a_2$, each coupled to a different transmission line, $TL_{1}$ and $TL_2$.
The lower resonator $a_1$ (green) connecting to the transmission line $TL_1$ denotes the ``bright" meta-atom that is excited directly by the incoming wave from $Port_1$, while the upper resonator $a_2$ (orange) connecting to $TL_2$ denotes the ``dark" meta-atom that can be excited by the ``bright" meta-atom. 
The lumped elements loaded on two meta-atoms can tune experimental parameters; for instance, the tunable resistors can change the dissipative loss of meta-atoms and the tunable capacitors can modulate the resonant frequency of each meta-atom. The element in red circle on ``bright" meta-atom depicts a varactor, which can induce the bistable state in the second experiment, but its equivalent capacitance is almost fixed throughout this work (see ``\textbf{Method}" for more details). As a consequence, we can use the following linear coupled mode theory to describe that system:
\begin{eqnarray}
%\label{eq:a1}
\frac{da_1}{dt}&=&(-i\omega_1-\gamma_1-\Gamma_1)a_1+i{\kappa}a_2+\sqrt{\gamma_1}S_{in}e^{-i{\omega}t},     \label{eq:a1}\\
%\end{equation}
%\begin{equation}
%\label{eq:a2}
\frac{da_2}{dt}&=&(-i\omega_2-\gamma_2-\Gamma_2)a_2+i{\kappa}a_1.
\label{eq:a2}
\end{eqnarray}
%where $a_{1,2}$ denotes the intra-cavity mode fields of the two meta-atoms; $\omega_{1,2}$ denote the resonant frequencies; $\gamma_{1,2}$ and $\Gamma_{1,2}$ denote the radiative losses and dissipative losses respectively; $\kappa$ denotes the near-filed coupling strength and $S_{in}$ is the incoming wave with time harmonic field $e^{-i{\omega}t}$.
To simplify the theoretical model, the resonant frequency of the two meta-atoms were set to $\omega_1=\omega_2=\omega_0$ via the tunable capacitor. In this basis, the effective Hamiltonian of the system can be written in a matrix format, as
\begin{equation}
\label{eq:M}
{M}=\left(
\begin{array}{cc}   
    \omega_0-i(\gamma_1+\Gamma_1) & -\kappa \\  
    -\kappa & \omega_0-i(\gamma_2+\Gamma_2) \\ 
  \end{array}
\right),
\end{equation}
of which the characteristic equation can be calculated from
$\vert{\omega}I-M\vert=0$. Consequently, the eigenfrequencies of modes formed by these coupled meta-atoms are obtained as
\begin{equation}
\label{eq:eigenfre}
\omega_{\pm}=\omega_0-iX\pm\sqrt{{\kappa}^2-Y^2},
\end{equation}
where $X=(\gamma_1+\Gamma_1+\gamma_2+\Gamma_2)/2$, $Y=(\gamma_1+\Gamma_1-\gamma_2-\Gamma_2)/2$. The evolution of eigenfrequenies is calculated in Fig.~\ref{fig:calculated re&im intensity}(a). The first panel depicts the real parts of  eigenfrequencies, which indicates that the two modes coalesce at an EP with increasing dissipative loss $\Gamma_2$ in the ``dark" meta-atom. After undergoing the EP, their imaginary parts are repelled, shown in the second panel, which results in an increasing imaginary part for one of the eigenfrequencies and a decreasing imaginary part for the other. As a consequence, one of the modes becomes less lossy, while the other becomes more lossy. 

To uncover clear physics of field intensity revival, we substitute $a_k=A_{k}e^{-i{\omega}t}$ and $\frac{da_k}{dt}=-i{\omega}A_{k}e^{-i{\omega}t}+\frac{dA_k}{dt}e^{-i{\omega}t}$ ($k=1,2$) in Eqs.~(\ref{eq:a1},\ref{eq:a2}), and we obtain
\begin{eqnarray}
\frac{dA_1}{dt}&=&(i\Delta_1-\gamma_1-\Gamma_1)A_1+i{\kappa}A_2+\sqrt{\gamma_1}S_{in},      \label{eq:dA1} \\
%\end{equation}
%\begin{equation}
\frac{dA_2}{dt}&=&(i\Delta_2-\gamma_2-\Gamma_2)A_2+i{\kappa}A_1,     
\label{eq:dA2}
\end{eqnarray}
where $\Delta_1=\omega-\omega_1$ and $\Delta_2=\omega-\omega_2$ are the detuning between the resonant frequency and the frequency of the incoming wave $S_{in}$. Further, we can solve Eqs.~(\ref{eq:dA1},\ref{eq:dA2}) at steady state and find the intra-cavity fields $A_1$ and $A_2$
\begin{eqnarray}
A_1&=&\frac{\textcolor[rgb]{0,0.5,1.0}{-}\sqrt{\gamma_1}(\gamma_2+\Gamma_2-i\Delta_0)}{{\kappa}^2+(\gamma_1+\Gamma_1-i\Delta_0)(\gamma_2+\Gamma_2-i\Delta_0)}S_{in},         \label{eq:A1}  \\
%\end{equation}
%\begin{equation}
A2&=&\frac{{i\kappa}\sqrt{\gamma_1}}{{\kappa}^2+(\gamma_1+\Gamma_1-i\Delta_0)(\gamma_2+\Gamma_2-i\Delta_0)}S_{in}.
\label{eq:A2}
\end{eqnarray}
Here, $\Delta_0=\Delta_{1,2}=\omega-\omega_0$ because the resonant frequencies of the two meta-atoms are set identical in our scheme. Based on Eq.~(\ref{eq:eigenfre}) and Eqs.~(\ref{eq:A1},\ref{eq:A2}) we can calculate the field intensity $I_{1,2}={\vert{A_{1,2}}\vert}^2$ at eigenfrequencies with the parameters fit from our system. Figure~\ref{fig:calculated re&im intensity}(b) shows evolution of field intensities (normalized by the total field intensity at $\Gamma_2=0$) at modes $\omega_{\pm}$. % in ``bright" meta-atom (green line), ``dark" meta-atom (orange line) and total intensity (blue line, $I_0=I_1+I_2$) with increasing dissipative loss $\Gamma_2$. 
At first, the field intensities in both meta-atom are nearly equal, and they both decreases with enlarging dissipative loss in the ``dark" meta-atom. When system is in the vicinity of the EP, the field intensity in ``bright" meta-atom recovers, which however does not occur in the ``dark'' one. After passing the EP, the field intensity revives completely even with enormous loss in the microwave photonic meta-molecule. Next, we are going to demonstrate such counterintuitive features in our experiments.

\begin{figure}
\includegraphics[width=1.0\linewidth]{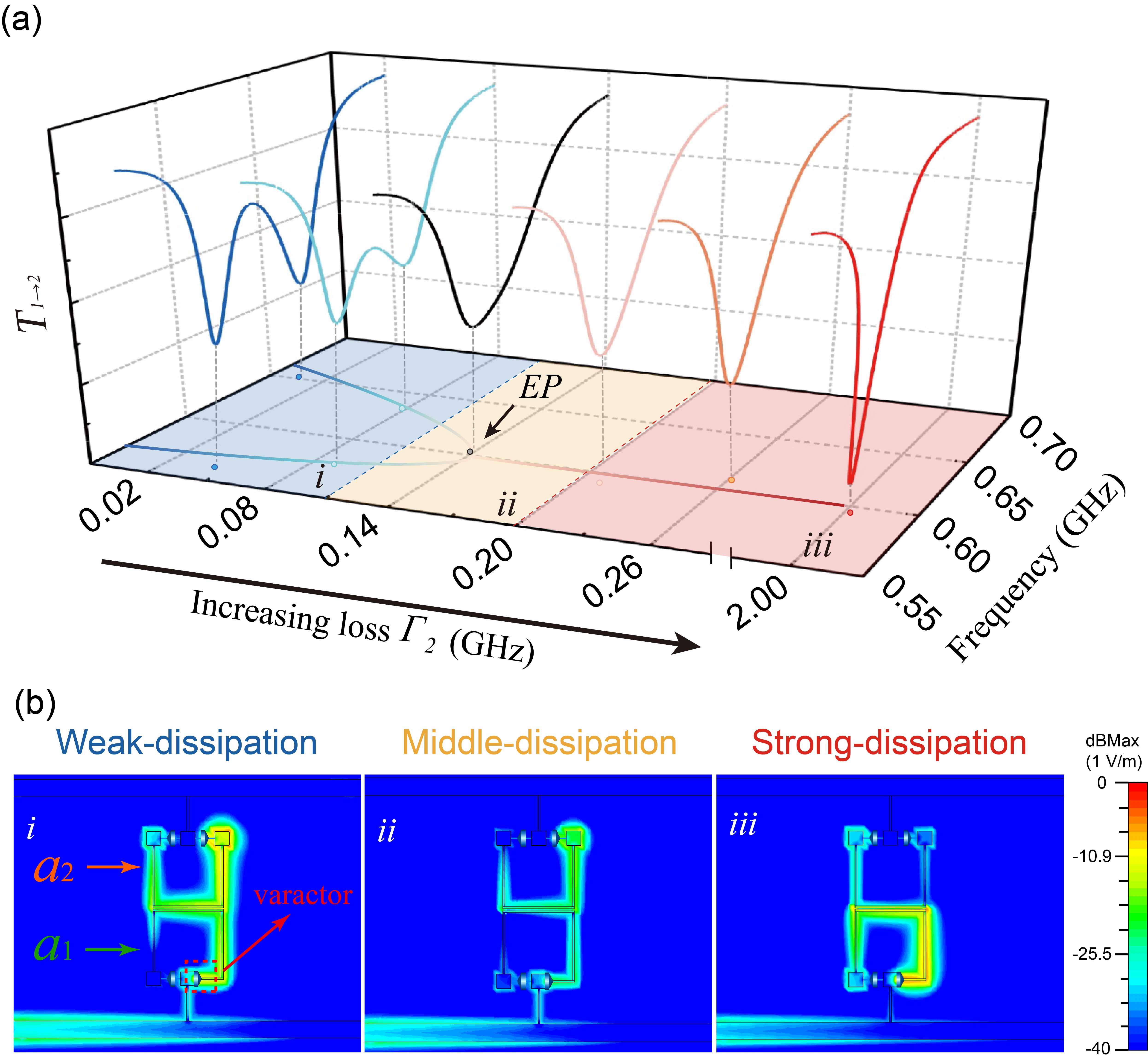}
\caption{\label{fig:s21&field}(Color online) (a) Measured (soild line) and calculated (dash line) transmission $T_{1\rightarrow2}$ (from $Port_1$ to $Port_2$) in three regimes with increasing dissipative loss $\Gamma_2$. $(i)$ weak-dissipation regime (blue), $(ii)$ middle-dissipation regime (yellow) and $(iii)$ strong-dissipation regime (red) are depicted in the bottom panel distinctly. Solid lines denote the calculated real parts of the eigenfrequencies and dots denote the experimentally measured resonant frequencies. (b) Corresponding simulated field intensity in ``Bright" ($a_1$) and ``Dark" ($a_2$) meta-atoms. Red dash circle on $a_1$ indicates the place where the varactor (or the capacitor) is loaded. The field intensity revival in ``bright" meta-atom $a_1$ is clear. }
\end{figure}

In the first set of experiments, we investigate the evolution of eigenfrequencies and transmission spectra $\mathit{T}_{1\rightarrow2}$ by continuously increasing $\Gamma_2$ (experimentally, increasing the resistor $R_2$). The transmittance $T_{1\rightarrow2}$ of the system from the input $Port_1$ to output $Port_2$ can be analytically obtained from Eqs.~(\ref{eq:A1}) as
\begin{eqnarray}
T_{1\rightarrow2}&=&{\vert{\frac{S_{in}-\sqrt{\gamma_1} A_1}{S_{in}}}\vert}^2={\vert{1+\frac{\gamma_1(\gamma_2+\Gamma_2-i\Delta_0)}{{\kappa}^2+(\gamma_1+\Gamma_1-i\Delta_0)(\gamma_2+\Gamma_2-i\Delta_0)}}\vert}^2
\label{eq:T12}
\end{eqnarray} 
As a consequences, we depicted both the theoretical and experimental results in Fig.~\ref{fig:s21&field}(a) and found them match very well. The coupling between two meta-atoms as well their losses leads to the formation of two eigenmodes with complex eigenfrequencies. At beginning, both resistors are set as $\mathit{R}_1\approx\mathit{R}_2\approx 0$ $\Omega$ such that $\Gamma_1\approx\Gamma_2\approx 0$ GHz, namely, the system is in the weak-dissipation regime $\left(\mathit{i}\right)$ $\vert\mathit{Y}\vert<\kappa$. In this regime, the eigenmodes have different resonant frequencies (mode splitting), which is determined by the value of real part of the square root in Eq.~(\ref{eq:eigenfre}). As we increase the dissipative loss $\Gamma_2$ in ``dark" meta-atom, the real part of the square root declines rapidly till the special EP, where $\left(\mathit{ii}\right)$  $\vert\mathit{Y}\vert=\kappa$ and the eigenmodes coalesce therein,
 where $\Gamma_2=0.14$ GHz. With further increasing of $\Gamma_2$, the system enters the strong-dissipation regime, quantified by $\left(\mathit{iii}\right)$ $\vert\mathit{Y}\vert>\kappa$ and the square root values as a pair of pure conjugated imaginary numbers, which leads to the conjugate eigenmodes with resonant frequencies of the same real value.  
The corresponding simulated field intensity distribution of three regimes are exemplified in Fig.~\ref{fig:s21&field}(b), which clearly demonstrate the revival of the field intensity on the ``bright'' meta-atom $a_1$.
%Here, we use  equivalent capacitor $\mathit{C}_0=2.65$ pf in both the ``bright" and ``dark'' meta-atoms. 
We set the varactor in the red dash circle in the following experiment to yield the bistable state. Remarkably, the revival of the field intensity, especially in the place where mounting the varactor, in ``bright" meta-atom can be observed directly. 

\begin{figure}
\includegraphics[width=0.95\linewidth]{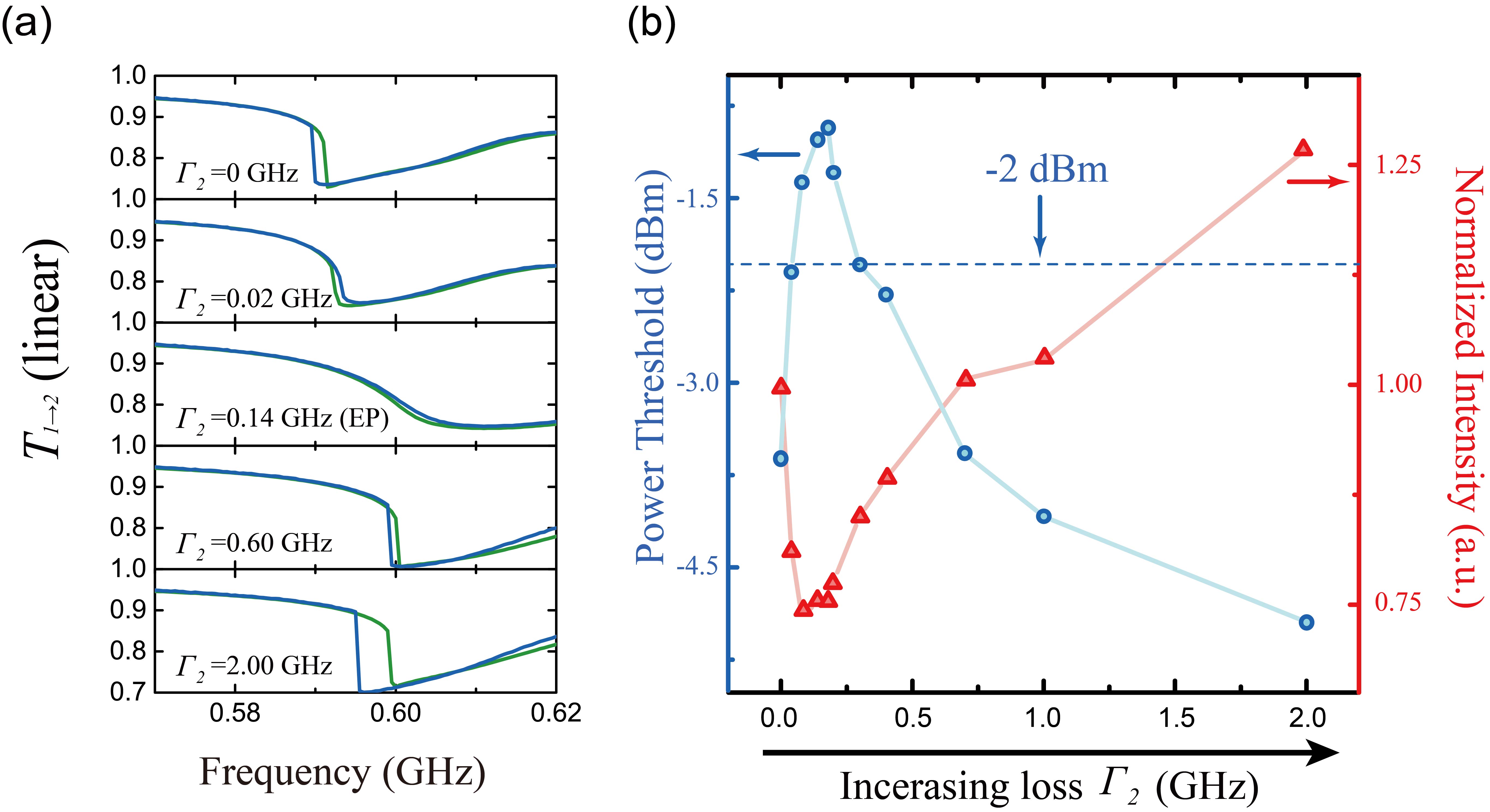}
\caption{\label{fig:Threshold}(Color online) (a) Measured $T_{1\rightarrow2}$ forward and backward frequency sweeping results with increasing dissipative loss $\Gamma_2$ and fixed power on $-2 dBm$. Note, we measured the mode $\omega_{-}$ because the bistable state induced by a varactor usually lowers the resonant frequency. (b) Loss-induced evolution of the power threshold (blue circle) and normalized field intensity (red triangle) in the meta-molecule. The blue dash line , with the same input power $-2 dBm$, indicates a non-monotone compare with the power threshold. }
\end{figure}

To further demonstrate the effect of revival on the bistable state we design the second set of experiments, where the capacitor in previous experiment is replaced by a nonlinear element, varactor, with equivalent capacitance. In this configuration, a bistable state can be stimulated with a strong field intensity above the threshold. This bistable behavior merely results from the nonlinearity of the varactor.  Fig.~\ref{fig:Threshold}(a) shows the effect of loss on frequency sweeping curves. The forward (blue line) and backward (green line) sweeping curves separate from each other (in the top panel), but then coalesce when the dissipative loss in ``dark" meta-atom approaches a critical value where the system is near the EP (in the middle panel). After further increasing loss to achieve a high level dissipation, the system passes through the EP, and  we can then observe the curves separate distinctly again (in the bottom panel). Note, the data are measured at the mode $\omega_{-}$ while at this mode one of the magnetic coupled meta-atoms has the nonlinear varactor. The bistable state does not result from the mode switching between $\omega_+$ and $\omega_-$, but is a consequence of the nonlinear response of the varactor. %where the bistable state emerge.  
Actually, we also find the similar result at power sweeping curves (not shown), which indicates that the bistable state indeed changes from attenuation to revival merely by increasing the loss. 

The feature that bistable state recovery in frequency sweeping curves is due to the loss-assisted field intensity revival in the vicinity of an EP. To support this result, we measure the field intensity in the ``bright" meta-atom and the power threshold of the bistable state, thanks to the varactor. As shown in Fig.~\ref{fig:Threshold}(b), %red triangle scatter line denotes the normalized intensity, and blue circle scatter line denotes the threshold. 
larger intensity, less threshold. When initially increasing loss,
the normalized intensity declines at first so that raises the threshold. When further increasing loss so that the system passes through the EP,  the increasing field intensity decreases the threshold. 
When we draw a dash line at our input power $-2 dBm$, the value of this dash line first becomes larger than the power threshold line, which means the bistable state is present; then with increasing dissipative loss $\Gamma_2$, the threshold becomes larger than the input power and  we find the bistable state vanished in this regime; while enlarging the loss, the threshold come down again and the bistable state revives. 
Moreover, we find the final threshold can be even smaller than the initial value (in the absence of loss) when the loss is large enough. The experimental results consist well with the theoretical results Eq.~(\ref{eq:A1}) and Fig.~\ref{Figure2}(b), which describe a loss-assisted bistable state (field intensity) revival in a non-Hermitian microwave photonic molecule with an EP.  The opposite evolution trends between the normalized intensity and bistable threshold tells us that the bistable loop is indeed induced by the nonlinear varactor, which is different from the mechanism of parameter loops in other topological works with linear elements~\cite{Doppler, Xu}.

\section{Methods}
\addcontentsline{toc}{Method}{Method}
The samples are all fabricated on copper-clad 0.787-mm thick Rogers RT5880 substrates using laser direct structuring technology (LPKF ProtoLaser 200). We use  equivalent capacitors $\mathit{C}_0=2.65$ pf in both the ``bright" and ``dark'' meta-atoms. For stimulating the bistable state, we put a varactor on $a_1$ whose capacitance $C_0=2.65$ pf but in the first set of experimental an equivalent capacitor was used instead of varactor. Two tunable resistors were loaded on resonators whose value can be changed from 0 to 100 $\Omega$. An additional tunable capacitor was set on $a_2$ to fix the resonant frequency error induced by manufacturing the sample. The distributive capacitance and inductance can be designed with different geometrical structure parameters. The two identical resonators are given as $\mathit{a}=10$ mm, $\mathit{d}=8$ mm, $\mathit{w}=0.2$ mm, $\mathit{s}=0.2$ mm, $\mathit{g}_1=0.8$ mm, $\mathit{g}_2=5$ mm. The four ports are fixed to transmission  lines by SMA linkers whose impedance equals $50$ $\Omega$. Transmission spectra $\mathit{T}_{1\rightarrow2}$ and $\mathit{T}_{1\rightarrow4}$ (the same as $\mathit{T}_{1\rightarrow3}$) are obtained directly from the microwave vector network analyzer (Agilent N5222A). In addition, a commercial software package (CST Microwave Studio) is used in designing the samples.

\section{Conclusions}
All of these observations are in contrast to what one would expect in conventional systems, where the higher loss usually causes the higher  bistable  threshold and thus detrimental to the bistable behavior. 
Now surprisingly, in the vicinity of an EP in our system, considerable loss will reversely reduce the bistable threshold and plays a constructive role to recover the bistable behavior.  Less loss is detrimental to optimize a bistable state system with low threshold. As is well-known, varactor can exhibit the bistable behavior once the strength of the field intensity exceeds some critical value.  Thus, the field intensity on the varactor of the resonator plays a key role. When the loss exceeds a critical value, the field strength of one eigenmode mostly locates in one of the subsystems with less loss (``bright" meta-atom in our present set-up) so as to avoid the dissipation. Thus, the total field can build up more strongly, which then helps the recovery of the bistable behavior. The observations may also put forward an alternative way to control and manifest the effect of loss in bistable state systems or other physical systems rely on an EP. We believe the results could pave the way for developing devices of switching, memory, sensor, as well as other logic operations, for all-optical communication and computing. 

\section*{Acknowledgments}
\addcontentsline{toc}{Acknowledgments}{Acknowledgments}
This research was supported by NKRDPC (Grant No. 2016YFA0301101), CNSF (Grants No. 11234010, No. 11204217, No.11674247 and No.61621001), the National Youth 1000 Talents Program in China, and the 985 startup grant (No. 205020516074) at Tongji University.

% 

%\bibliography{Reference}% Produces the bibliography via BibTeX.

\section*{References}
\addcontentsline{toc}{References}{References}
\begin{thebibliography}{34}%
\makeatletter
\providecommand \@ifxundefined [1]{%
 \@ifx{#1\undefined}
}%
\providecommand \@ifnum [1]{%
 \ifnum #1\expandafter \@firstoftwo
 \else \expandafter \@secondoftwo
 \fi
}%
\providecommand \@ifx [1]{%
 \ifx #1\expandafter \@firstoftwo
 \else \expandafter \@secondoftwo
 \fi
}%
\providecommand \natexlab [1]{#1}%
\providecommand \enquote  [1]{``#1''}%
\providecommand \bibnamefont  [1]{#1}%
\providecommand \bibfnamefont [1]{#1}%
\providecommand \citenamefont [1]{#1}%
\providecommand \href@noop [0]{\@secondoftwo}%
\providecommand \href [0]{\begingroup \@sanitize@url \@href}%
\providecommand \@href[1]{\@@startlink{#1}\@@href}%
\providecommand \@@href[1]{\endgroup#1\@@endlink}%
\providecommand \@sanitize@url [0]{\catcode `\\12\catcode `\$12\catcode
  `\&12\catcode `\#12\catcode `\^12\catcode `\_12\catcode `\%12\relax}%
\providecommand \@@startlink[1]{}%
\providecommand \@@endlink[0]{}%
\providecommand \url  [0]{\begingroup\@sanitize@url \@url }%
\providecommand \@url [1]{\endgroup\@href {#1}{\urlprefix }}%
\providecommand \urlprefix  [0]{URL }%
\providecommand \Eprint [0]{\href }%
\providecommand \doibase [0]{http://dx.doi.org/}%
\providecommand \selectlanguage [0]{\@gobble}%
\providecommand \bibinfo  [0]{\@secondoftwo}%
\providecommand \bibfield  [0]{\@secondoftwo}%
\providecommand \translation [1]{[#1]}%
\providecommand \BibitemOpen [0]{}%
\providecommand \bibitemStop [0]{}%
\providecommand \bibitemNoStop [0]{.\EOS\space}%
\providecommand \EOS [0]{\spacefactor3000\relax}%
\providecommand \BibitemShut  [1]{\csname bibitem#1\endcsname}%
\let\auto@bib@innerbib\@empty
%</preamble>
\bibitem{bender2007making}%
  \BibitemOpen
  \bibfield  {author} {\bibinfo {author} {\bibfnamefont {C.~M.}\ \bibnamefont
  {Bender}},\ }\href@noop {} {\bibfield  {journal} {\bibinfo  {journal}
  {\textit{Rep. Prog. Phys}}\ }\textbf {\bibinfo {volume} {70}},\
  \bibinfo {pages} {947} (\bibinfo {year} {2007})}\BibitemShut {NoStop}%
\bibitem{kato1976perturbation}%
  T. Kato, "Perturbation theory of linear operators", \textit{Springer-Verlag Berling Heridelberg}, (1976).
\bibitem {heiss2012physics}%
  \BibitemOpen
  \bibfield  {author} {\bibinfo {author} {\bibfnamefont {W.}~\bibnamefont
  {Heiss}},\ }\href@noop {} {\bibfield  {journal} {\bibinfo  {journal} {\textit{J. Phys. A: Math. Theor}}\ }\textbf {\bibinfo {volume}
  {45}},\ \bibinfo {pages} {444016} (\bibinfo {year} {2012})}\BibitemShut
  {NoStop}%
\bibitem {dembowski2004encircling}%
  \BibitemOpen
  \bibfield  {author} {\bibinfo {author} {\bibfnamefont {C.}~\bibnamefont
  {Dembowski}}, \bibinfo {author} {\bibfnamefont {B.}~\bibnamefont {Dietz}},
  \bibinfo {author} {\bibfnamefont {H.-D.}\ \bibnamefont {Gr{\"a}f}}, \bibinfo
  {author} {\bibfnamefont {H.}~\bibnamefont {Harney}}, \bibinfo {author}
  {\bibfnamefont {A.}~\bibnamefont {Heine}}, \bibinfo {author} {\bibfnamefont
  {W.}~\bibnamefont {Heiss}}, \ and\ \bibinfo {author} {\bibfnamefont
  {A.}~\bibnamefont {Richter}},\ }\href@noop {} {\bibfield  {journal} {\bibinfo
   {journal} {\textit{Phys. Rev. E}}\ }\textbf {\bibinfo {volume} {69}},\ \bibinfo
  {pages} {056216} (\bibinfo {year} {2004})}\BibitemShut {NoStop}%
\bibitem{yannopapas2013spontaneous}%
  \BibitemOpen
  \bibfield  {author} {\bibinfo {author} {\bibfnamefont {V.}\ \bibnamefont
  {Yannopapas}},\ }\href@noop {} {\bibfield  {journal} {\bibinfo  {journal}
  {\textit{Phys. Rev. A}}\ }\textbf {\bibinfo {volume} {89}},\ \bibinfo
  {pages} {013808}(\bibinfo {year} {2014})}\BibitemShut {NoStop}%\
\bibitem{yannopapas2012non}%
  \BibitemOpen
  \bibfield  {author} {\bibinfo {author} {\bibfnamefont {V.}\ \bibnamefont
  {Yannopapas}},\ }\href@noop {} {\bibfield  {journal} {\bibinfo  {journal}
  {\textit{J. Opt.}}\ }\textbf {\bibinfo {volume} {14}},\ \bibinfo
  {pages} {085105}(\bibinfo {year} {2012})}\BibitemShut {NoStop}%\
\bibitem{yannopapas2012topological}%
  \BibitemOpen
  \bibfield  {author} {\bibinfo {author} {\bibfnamefont {V.}\ \bibnamefont
  {Yannopapas}},\ }\href@noop {} {\bibfield  {journal} {\bibinfo  {journal}
  {\textit{New J. Phys}}\ }\textbf {\bibinfo {volume} {14}},\ \bibinfo
  {pages} {113017}(\bibinfo {year} {2012})}\BibitemShut {NoStop}%\
\bibitem {guo2009observation}%
  \BibitemOpen
  \bibfield  {author} {\bibinfo {author} {\bibfnamefont {A.}~\bibnamefont
  {Guo}}, \bibinfo {author} {\bibfnamefont {G.}~\bibnamefont {Salamo}},
  \bibinfo {author} {\bibfnamefont {D.}~\bibnamefont {Duchesne}}, \bibinfo
  {author} {\bibfnamefont {R.}~\bibnamefont {Morandotti}}, \bibinfo {author}
  {\bibfnamefont {M.}~\bibnamefont {Volatier-Ravat}}, \bibinfo {author}
  {\bibfnamefont {V.}~\bibnamefont {Aimez}}, \bibinfo {author} {\bibfnamefont
  {G.}~\bibnamefont {Siviloglou}}, \ and\ \bibinfo {author} {\bibfnamefont
  {D.}~\bibnamefont {Christodoulides}},\ }\href@noop {} {\bibfield  {journal}
  {\bibinfo  {journal} {\textit{Phys. Rev. Lett.}}\ }\textbf {\bibinfo {volume}
  {103}},\ \bibinfo {pages} {093902} (\bibinfo {year} {2009})}\BibitemShut
  {NoStop}%
\bibitem {dembowski2001experimental}%
  \BibitemOpen
  \bibfield  {author} {\bibinfo {author} {\bibfnamefont {C.}~\bibnamefont
  {Dembowski}}, \bibinfo {author} {\bibfnamefont {H.-D.}\ \bibnamefont
  {Gr{\"a}f}}, \bibinfo {author} {\bibfnamefont {H.}~\bibnamefont {Harney}},
  \bibinfo {author} {\bibfnamefont {A.}~\bibnamefont {Heine}}, \bibinfo
  {author} {\bibfnamefont {W.}~\bibnamefont {Heiss}}, \bibinfo {author}
  {\bibfnamefont {H.}~\bibnamefont {Rehfeld}}, \ and\ \bibinfo {author}
  {\bibfnamefont {A.}~\bibnamefont {Richter}},\ }\href@noop {} {\bibfield
  {journal} {\bibinfo  {journal} {\textit{Phys. Rev. Lett.}}\ }\textbf {\bibinfo
  {volume} {86}},\ \bibinfo {pages} {787} (\bibinfo {year} {2001})}\BibitemShut
  {NoStop}%
\bibitem {cerjan2016eigenvalue}%
  \BibitemOpen
  \bibfield  {author} {\bibinfo {author} {\bibfnamefont {Alexander}\ \bibnamefont
  {Cerjan}},\ and\ {\bibfnamefont {Shanhui}\ \bibnamefont
  {Fan}} }\href@noop {} {\bibfield  {journal} {\bibinfo  {journal}
  {\textit{Phys. Rev. A}}\ }\textbf {\bibinfo {volume} {94}},\
  \bibinfo {pages} {033857} (\bibinfo {year} {2016})}\BibitemShut {NoStop}%
\bibitem{gao2015observation}%
  \BibitemOpen
  \bibfield  {author} {\bibinfo {author} {\bibfnamefont {T.}~\bibnamefont
  {Gao}}, \bibinfo {author} {\bibfnamefont {E.}~\bibnamefont {Estrecho}},
  \bibinfo {author} {\bibfnamefont {K.}~\bibnamefont {Bliokh}}, \bibinfo
  {author} {\bibfnamefont {T.}~\bibnamefont {Liew}}, \bibinfo {author}
  {\bibfnamefont {M.}~\bibnamefont {Fraser}}, \bibinfo {author} {\bibfnamefont
  {S.}~\bibnamefont {Brodbeck}}, \bibinfo {author} {\bibfnamefont
  {M.}~\bibnamefont {Kamp}}, \bibinfo {author} {\bibfnamefont {C.}~\bibnamefont
  {Schneider}}, \bibinfo {author} {\bibfnamefont {S.}~\bibnamefont
  {H{\"o}fling}}, \bibinfo {author} {\bibfnamefont {Y.}~\bibnamefont
  {Yamamoto}},\ }\href@noop {} {\bibfield  {journal} {\bibinfo
   {journal} {\textit{Nature}}\ }\textbf {\bibinfo {volume} {526}},\ \bibinfo {pages}
  {554}(\bibinfo {year}
  {2015})}\BibitemShut {NoStop}%
\bibitem {peng2014loss}%
  \BibitemOpen
  \bibfield  {author} {\bibinfo {author} {\bibfnamefont {B.}~\bibnamefont
  {Peng}}, \bibinfo {author} {\bibfnamefont {{\c{S}}.}~\bibnamefont
  {{\"O}zdemir}}, \bibinfo {author} {\bibfnamefont {S.}~\bibnamefont {Rotter}},
  \bibinfo {author} {\bibfnamefont {H.}~\bibnamefont {Yilmaz}}, \bibinfo
  {author} {\bibfnamefont {M.}~\bibnamefont {Liertzer}}, \bibinfo {author}
  {\bibfnamefont {F.}~\bibnamefont {Monifi}}, \bibinfo {author} {\bibfnamefont
  {C.}~\bibnamefont {Bender}}, \bibinfo {author} {\bibfnamefont
  {F.}~\bibnamefont {Nori}}, \ and\ \bibinfo {author} {\bibfnamefont
  {L.}~\bibnamefont {Yang}},\ }\href@noop {} {\bibfield  {journal} {\bibinfo
  {journal} {\textit{Science}}\ }\textbf {\bibinfo {volume} {346}},\ \bibinfo {pages}
  {328} (\bibinfo {year} {2014})}\BibitemShut {NoStop}%
  
\bibitem {brandstetter2014reversing}%
  \BibitemOpen
  \bibfield  {author} {\bibinfo {author} {\bibfnamefont {M.}~\bibnamefont
  {Brandstetter}}, \bibinfo {author} {\bibfnamefont {M.}~\bibnamefont
  {Liertzer}}, \bibinfo {author} {\bibfnamefont {C.}~\bibnamefont {Deutsch}},
  \bibinfo {author} {\bibfnamefont {P.}~\bibnamefont {Klang}}, \bibinfo
  {author} {\bibfnamefont {J.}~\bibnamefont {Sch{\"o}berl}}, \bibinfo {author}
  {\bibfnamefont {H.}~\bibnamefont {T{\"u}reci}}, \bibinfo {author}
  {\bibfnamefont {G.}~\bibnamefont {Strasser}}, \bibinfo {author}
  {\bibfnamefont {K.}~\bibnamefont {Unterrainer}}, \ and\ \bibinfo {author}
  {\bibfnamefont {S.}~\bibnamefont {Rotter}},\ }\href@noop {} {\bibfield
  {journal} {\bibinfo  {journal} {\textit{Nat. commun.}}\ }\textbf {\bibinfo
  {volume} {5}},\ \bibinfo {pages} {4034} (\bibinfo {year} {2014})}\BibitemShut {NoStop}%
\bibitem{Chitsazi2014Experimental}%
  \BibitemOpen
  \bibfield  {author} {\bibinfo {author} {\bibfnamefont {M.}~\bibnamefont
  {Chitsazi}}, \bibinfo {author} {\bibfnamefont {J.}~\bibnamefont {Schindler}}, \bibinfo
  {author} {\bibfnamefont {H.}\ \bibnamefont {Ramezani}}, \bibinfo
  {author} {\bibfnamefont {F. M.}\ \bibnamefont {Ellis}}, \ and\ \bibinfo
  {author} {\bibfnamefont {T.}~\bibnamefont {Kottos}},\ }\href@noop {} {\bibfield
   {journal} {\bibinfo  {journal} {\textit{Phys. Rev. A}}\ }\textbf {\bibinfo
  {volume} {89}},\ \bibinfo {pages} {123-127} (\bibinfo {year}
  {2014})}\BibitemShut {NoStop}%
\bibitem {miri2012large}%
  \BibitemOpen
  \bibfield  {author} {\bibinfo {author} {\bibfnamefont {M. A.}~\bibnamefont
  {Miri}}, \bibinfo {author} {\bibfnamefont {P.}~\bibnamefont
  {Likamwa}}, \ and\ \bibinfo {author} {\bibfnamefont
  {D. N.}~\bibnamefont {Christodoulides}},\ }\href@noop {} {\bibfield  {journal} {\bibinfo
  {journal} {\textit{Opt. Lett.}}\ }\textbf {\bibinfo {volume} {37}},\ \bibinfo {pages}{764-766} (\bibinfo {year} {2012})}\BibitemShut {NoStop}%
\bibitem {feng2014single}%
  \BibitemOpen
  \bibfield  {author} {\bibinfo {author} {\bibfnamefont {L.}~\bibnamefont
  {Feng}}, \bibinfo {author} {\bibfnamefont {Z. J.}~\bibnamefont
  {Wong}}, \bibinfo {author} {\bibfnamefont {R. M.}~\bibnamefont
  {Ma}}, \bibinfo {author} {\bibfnamefont
  {Y.}~\bibnamefont {Wang}}, \ and\ \bibinfo {author} {\bibfnamefont
  {X.}~\bibnamefont {Zhang}},\ }\href@noop {} {\bibfield  {journal} {\bibinfo
  {journal} {\textit{Science}}\ }\textbf {\bibinfo {volume} {346}},\ \bibinfo {pages}
  {972-975} (\bibinfo {year} {2014})}\BibitemShut {NoStop}%
\bibitem {hodaei2014parity}%
  \BibitemOpen
  \bibfield  {author} {\bibinfo {author} {\bibfnamefont {H.}~\bibnamefont
  {Hodaei}}, \bibinfo {author} {\bibfnamefont {M. A.}~\bibnamefont
  {Miri}}, \bibinfo {author} {\bibfnamefont {M.}~\bibnamefont
  {Heinrich}}, \bibinfo {author} {\bibfnamefont
  {D. N.}~\bibnamefont {Christodoulides}}, \ and\ \bibinfo {author} {\bibfnamefont
  {M.}~\bibnamefont {Khajavikhan}},\ }\href@noop {} {\bibfield  {journal} {\bibinfo
  {journal} {\textit{Science}}\ }\textbf {\bibinfo {volume} {346}},\ \bibinfo {pages}
  {975-978} (\bibinfo {year} {2014})}\BibitemShut {NoStop}% 
\bibitem {gentry2014dark}%
  \BibitemOpen
  \bibfield  {author} {\bibinfo {author} {\bibfnamefont {C. M.}~\bibnamefont
  {Gentry}}, \ and\ \bibinfo {author} {\bibfnamefont
  {M. A.}~\bibnamefont {Popović}},\ }\href@noop {} {\bibfield  {journal} {\bibinfo
  {journal} {\textit{Opt. Lett.}}\ }\textbf {\bibinfo {volume} {39}},\ \bibinfo {pages}
  {4136-4139} (\bibinfo {year} {2014})}\BibitemShut {NoStop}% 
\bibitem {dembowski2003observation}%
  \BibitemOpen
  \bibfield  {author} {\bibinfo {author} {\bibfnamefont {C.}~\bibnamefont
  {Dembowski}}, \bibinfo {author} {\bibfnamefont {B.}~\bibnamefont {Dietz}},
  \bibinfo {author} {\bibfnamefont {H.-D.}\ \bibnamefont {Gr{\"a}f}}, \bibinfo
  {author} {\bibfnamefont {H.}~\bibnamefont {Harney}}, \bibinfo {author}
  {\bibfnamefont {A.}~\bibnamefont {Heine}}, \bibinfo {author} {\bibfnamefont
  {W.}~\bibnamefont {Heiss}}, \ and\ \bibinfo {author} {\bibfnamefont
  {A.}~\bibnamefont {Richter}},\ }\href@noop {} {\bibfield  {journal} {\bibinfo
   {journal} {\textit{Phys. Rev. Lett.}}\ }\textbf {\bibinfo {volume} {90}},\
  \bibinfo {pages} {034101} (\bibinfo {year} {2003})}\BibitemShut {NoStop}%
\bibitem {heiss2000repulsion}%
  \BibitemOpen
  \bibfield  {author} {\bibinfo {author} {\bibfnamefont {W.}~\bibnamefont
  {Heiss}},\ }\href@noop {} {\bibfield  {journal} {\bibinfo  {journal}
  {\textit{Phys. Rev. E}}\ }\textbf {\bibinfo {volume} {61}},\ \bibinfo {pages}
  {929} (\bibinfo {year} {2000})}\BibitemShut {NoStop}%
\bibitem {bender1998real}%
  \BibitemOpen
  \bibfield  {author} {\bibinfo {author} {\bibfnamefont {C.~M.}\ \bibnamefont
  {Bender}}\ and\ \bibinfo {author} {\bibfnamefont {S.}~\bibnamefont
  {Boettcher}},\ }\href@noop {} {\bibfield  {journal} {\bibinfo  {journal}
  {\textit{Phys. Rev. Lett.}}\ }\textbf {\bibinfo {volume} {80}},\ \bibinfo
  {pages} {5243} (\bibinfo {year} {1998})}\BibitemShut {NoStop}%
\bibitem{el2007theory}%
  \BibitemOpen
  \bibfield  {author} {\bibinfo {author} {\bibfnamefont {R.}~\bibnamefont
  {El-Ganainy}}, \bibinfo {author} {\bibfnamefont {K.}~\bibnamefont {Makris}},
  \bibinfo {author} {\bibfnamefont {D.}~\bibnamefont {Christodoulides}}, \ and\
  \bibinfo {author} {\bibfnamefont {Z.~H.}\ \bibnamefont {Musslimani}},\
  }\href@noop {} {\bibfield  {journal} {\bibinfo  {journal} {\textit{Opt. Lett.}}\
  }\textbf {\bibinfo {volume} {32}},\ \bibinfo {pages} {2632} (\bibinfo {year}
  {2007})}\BibitemShut {NoStop}%
\bibitem{ruter2010observation}%
  \BibitemOpen
  \bibfield  {author} {\bibinfo {author} {\bibfnamefont {C.~E.}\ \bibnamefont
  {R{\"u}ter}}, \bibinfo {author} {\bibfnamefont {K.~G.}\ \bibnamefont
  {Makris}}, \bibinfo {author} {\bibfnamefont {R.}~\bibnamefont {El-Ganainy}},
  \bibinfo {author} {\bibfnamefont {D.~N.}\ \bibnamefont {Christodoulides}},
  \bibinfo {author} {\bibfnamefont {M.}~\bibnamefont {Segev}}, \ and\ \bibinfo
  {author} {\bibfnamefont {D.}~\bibnamefont {Kip}},\ }\href@noop {} {\bibfield
  {journal} {\bibinfo  {journal} {\textit{Nat. Phys.}}\ }\textbf {\bibinfo {volume}
  {6}},\ \bibinfo {pages} {192} (\bibinfo {year} {2010})}\BibitemShut {NoStop}%
\bibitem {klaiman2008visualization}%
  \BibitemOpen
  \bibfield  {author} {\bibinfo {author} {\bibfnamefont {S.}~\bibnamefont
  {Klaiman}}, \bibinfo {author} {\bibfnamefont {U.}~\bibnamefont
  {G{\"u}nther}}, \ and\ \bibinfo {author} {\bibfnamefont {N.}~\bibnamefont
  {Moiseyev}},\ }\href@noop {} {\bibfield  {journal} {\bibinfo  {journal}
  {\textit{Phys. Rev. Lett.}}\ }\textbf {\bibinfo {volume} {101}},\ \bibinfo
  {pages} {080402} (\bibinfo {year} {2008})}\BibitemShut {NoStop}%
\bibitem {sun2014experimental}%
  \BibitemOpen
  \bibfield  {author} {\bibinfo {author} {\bibfnamefont {Y.}~\bibnamefont
  {Sun}}, \bibinfo {author} {\bibfnamefont {W.}~\bibnamefont {Tan}}, \bibinfo
  {author} {\bibfnamefont {H.-Q.}\ \bibnamefont {Li}}, \bibinfo {author}
  {\bibfnamefont {J.}~\bibnamefont {Li}}, \ and\ \bibinfo {author}
  {\bibfnamefont {H.}~\bibnamefont {Chen}},\ }\href@noop {} {\bibfield
  {journal} {\bibinfo  {journal} {\textit{Phys. Rev. Lett.}}\ }\textbf {\bibinfo
  {volume} {112}},\ \bibinfo {pages} {143903} (\bibinfo {year}
  {2014})}\BibitemShut {NoStop}% 
\bibitem {Verslegers2012From}%
  \BibitemOpen
  \bibfield  {author} {\bibinfo {author} {\bibfnamefont {Lieven}~\bibnamefont
  {Verslegers}}, \bibinfo {author} {\bibfnamefont {Zongfu}~\bibnamefont {Yu}}, \bibinfo
  {author} {\bibfnamefont {Zhichao}\ \bibnamefont {Ruan}}, \bibinfo {author}
  {\bibfnamefont {Peter B.}~\bibnamefont {Catrysse}}, \ and\ \bibinfo {author}
  {\bibfnamefont {Shanhui}~\bibnamefont {Fan}},\ }\href@noop {} {\bibfield
  {journal} {\bibinfo  {journal} {\textit{Phys. Rev. Lett.}}\ }\textbf {\bibinfo
  {volume} {108}},\ \bibinfo {pages} {083902} (\bibinfo {year}
  {2012})}\BibitemShut {NoStop}%
\bibitem {fan2014tunable}%
  \BibitemOpen
  \bibfield  {author} {\bibinfo {author} {\bibfnamefont {Y.}~\bibnamefont
  {Fan}}, \bibinfo {author} {\bibfnamefont {F.}~\bibnamefont {Zhang}}, \bibinfo
  {author} {\bibfnamefont {Q.}~\bibnamefont {Zhao}}, \bibinfo {author}
  {\bibfnamefont {Z.}~\bibnamefont {Wei}}, \ and\ \bibinfo {author}
  {\bibfnamefont {H.}~\bibnamefont {Li}},\ }\href@noop {} {\bibfield  {journal}
  {\bibinfo  {journal} {\textit{Opt. Lett.}}\ }\textbf {\bibinfo {volume} {39}},\
  \bibinfo {pages} {6269} (\bibinfo {year} {2014})}\BibitemShut {NoStop}%
\bibitem {regensburger2012parity}%
  \BibitemOpen
  \bibfield  {author} {\bibinfo {author} {\bibfnamefont {A.}~\bibnamefont
  {Regensburger}}, \bibinfo {author} {\bibfnamefont {C.}~\bibnamefont
  {Bersch}}, \bibinfo {author} {\bibfnamefont {M. A.}\ \bibnamefont {Miri}},
  \bibinfo {author} {\bibfnamefont {G.}~\bibnamefont {Onishchukov}}, \bibinfo
  {author} {\bibfnamefont {D.~N.}\ \bibnamefont {Christodoulides}}, \ and\
  \bibinfo {author} {\bibfnamefont {U.}~\bibnamefont {Peschel}},\ }\href@noop
  {} {\bibfield  {journal} {\bibinfo  {journal} {\textit{Nature}\ }}\textbf {\bibinfo
  {volume} {488}},\ \bibinfo {pages} {167} (\bibinfo {year}
  {2012})}\BibitemShut {NoStop}%
\bibitem {bliokh2008coupling}%
  \BibitemOpen
  \bibfield  {author} {\bibinfo {author} {\bibfnamefont {K.~Y.}\ \bibnamefont
  {Bliokh}}, \bibinfo {author} {\bibfnamefont {Y.~P.}\ \bibnamefont {Bliokh}},
  \bibinfo {author} {\bibfnamefont {V.}~\bibnamefont {Freilikher}}, \bibinfo
  {author} {\bibfnamefont {A.}~\bibnamefont {Genack}}, \ and\ \bibinfo {author}
  {\bibfnamefont {P.}~\bibnamefont {Sebbah}},\ }\href@noop {} {\bibfield
  {journal} {\bibinfo  {journal} {\textit{Phys. Rev. Lett.}}\ }\textbf {\bibinfo
  {volume} {101}},\ \bibinfo {pages} {133901} (\bibinfo {year}
  {2008})}\BibitemShut {NoStop}%
\bibitem {cao2015dielectric}%
  \BibitemOpen
  \bibfield  {author} {\bibinfo {author} {\bibfnamefont {H.}~\bibnamefont
  {Cao}}\ and\ \bibinfo {author} {\bibfnamefont {J.}~\bibnamefont {Wiersig}},\
  }\href@noop {} {\bibfield  {journal} {\bibinfo  {journal} {\textit{Rev. Mod. Phys.}}\ }\textbf {\bibinfo {volume} {87}},\ \bibinfo {pages} {61} (\bibinfo
  {year} {2015})}\BibitemShut {NoStop}%
\bibitem {choi2010quasieigenstate}%
  \BibitemOpen
  \bibfield  {author} {\bibinfo {author} {\bibfnamefont {Y.}~\bibnamefont
  {Choi}}, \bibinfo {author} {\bibfnamefont {S.}~\bibnamefont {Kang}}, \bibinfo
  {author} {\bibfnamefont {S.}~\bibnamefont {Lim}}, \bibinfo {author}
  {\bibfnamefont {W.}~\bibnamefont {Kim}}, \bibinfo {author} {\bibfnamefont
  {J.-R.}\ \bibnamefont {Kim}}, \bibinfo {author} {\bibfnamefont {J.-H.}\
  \bibnamefont {Lee}}, \ and\ \bibinfo {author} {\bibfnamefont
  {K.}~\bibnamefont {An}},\ }\href@noop {} {\bibfield  {journal} {\bibinfo
  {journal} {\textit{Phys. Rev. Lett.}}\ }\textbf {\bibinfo {volume} {104}},\
  \bibinfo {pages} {153601} (\bibinfo {year} {2010})}\BibitemShut {NoStop}%
\bibitem {milner2001optical}%
  \BibitemOpen
  \bibfield  {author} {\bibinfo {author} {\bibfnamefont {V.}~\bibnamefont
  {Milner}}, \bibinfo {author} {\bibfnamefont {J.}~\bibnamefont {Hanssen}},
  \bibinfo {author} {\bibfnamefont {W.}~\bibnamefont {Campbell}}, \ and\
  \bibinfo {author} {\bibfnamefont {M.}~\bibnamefont {Raizen}},\ }\href@noop {}
  {\bibfield  {journal} {\bibinfo  {journal} {\textit{Phys. Rev. Lett.}}\
  }\textbf {\bibinfo {volume} {86}},\ \bibinfo {pages} {1514} (\bibinfo {year}
  {2001})}\BibitemShut {NoStop}%
\bibitem {kaplan2001observation}%
  \BibitemOpen
  \bibfield  {author} {\bibinfo {author} {\bibfnamefont {A.}~\bibnamefont
  {Kaplan}}, \bibinfo {author} {\bibfnamefont {N.}~\bibnamefont {Friedman}},
  \bibinfo {author} {\bibfnamefont {M.}~\bibnamefont {Andersen}}, \ and\
  \bibinfo {author} {\bibfnamefont {N.}~\bibnamefont {Davidson}},\ }\href@noop
  {} {\bibfield  {journal} {\bibinfo  {journal} {\textit{Phys. Rev. Lett.}}\
  }\textbf {\bibinfo {volume} {87}},\ \bibinfo {pages} {274101} (\bibinfo
  {year} {2001})}\BibitemShut {NoStop}%
\bibitem {akis1997wave}%
  \BibitemOpen
  \bibfield  {author} {\bibinfo {author} {\bibfnamefont {R.}~\bibnamefont
  {Akis}}, \bibinfo {author} {\bibfnamefont {D.}~\bibnamefont {Ferry}}, \ and\
  \bibinfo {author} {\bibfnamefont {J.}~\bibnamefont {Bird}},\ }\href@noop {}
  {\bibfield  {journal} {\bibinfo  {journal} {\textit{Phys. Rev. Lett.}}\
  }\textbf {\bibinfo {volume} {79}},\ \bibinfo {pages} {123} (\bibinfo {year}
  {1997})}\BibitemShut {NoStop}%
\bibitem{ponomarenko2008chaotic}%
  \BibitemOpen
  \bibfield  {author} {\bibinfo {author} {\bibfnamefont {L.}~\bibnamefont
  {Ponomarenko}}, \bibinfo {author} {\bibfnamefont {F.}~\bibnamefont
  {Schedin}}, \bibinfo {author} {\bibfnamefont {M.}~\bibnamefont {Katsnelson}},
  \bibinfo {author} {\bibfnamefont {R.}~\bibnamefont {Yang}}, \bibinfo {author}
  {\bibfnamefont {E.}~\bibnamefont {Hill}}, \bibinfo {author} {\bibfnamefont
  {K.}~\bibnamefont {Novoselov}}, \ and\ \bibinfo {author} {\bibfnamefont
  {A.}~\bibnamefont {Geim}},\ }\href@noop {} {\bibfield  {journal} {\bibinfo
  {journal} {\textit{Science}}\ }\textbf {\bibinfo {volume} {320}},\ \bibinfo {pages}
  {356} (\bibinfo {year} {2008})}\BibitemShut {NoStop}%
\bibitem {tan2014manipulating}%
  \BibitemOpen
  \bibfield  {author} {\bibinfo {author} {\bibfnamefont {W.}~\bibnamefont
  {Tan}}, \bibinfo {author} {\bibfnamefont {Y.}~\bibnamefont {Sun}}, \bibinfo
  {author} {\bibfnamefont {Z.-G.}\ \bibnamefont {Wang}}, \ and\ \bibinfo
  {author} {\bibfnamefont {H.}~\bibnamefont {Chen}},\ }\href@noop {} {\bibfield
   {journal} {\bibinfo  {journal} {\textit{Appl. Phys. Lett.}}\ }\textbf {\bibinfo
  {volume} {104}},\ \bibinfo {pages} {091107} (\bibinfo {year}
  {2014})}\BibitemShut {NoStop}%
\bibitem {liertzer2012pump}%
  \BibitemOpen
  \bibfield  {author} {\bibinfo {author} {\bibfnamefont {M.}~\bibnamefont
  {Liertzer}}, \bibinfo {author} {\bibfnamefont {L.}~\bibnamefont {Ge}},
  \bibinfo {author} {\bibfnamefont {A.}~\bibnamefont {Cerjan}}, \bibinfo
  {author} {\bibfnamefont {A.}~\bibnamefont {Stone}}, \bibinfo {author}
  {\bibfnamefont {H.}~\bibnamefont {T{\"u}reci}}, \ and\ \bibinfo {author}
  {\bibfnamefont {S.}~\bibnamefont {Rotter}},\ }\href@noop {} {\bibfield
  {journal} {\bibinfo  {journal} {\textit{Phys. Rev. Lett.}}\ }\textbf {\bibinfo
  {volume} {108}},\ \bibinfo {pages} {173901} (\bibinfo {year}
  {2012})}\BibitemShut {NoStop}%
\bibitem {wenzel1996mechanisms}%
  \BibitemOpen
  \bibfield  {author} {\bibinfo {author} {\bibfnamefont {H.}~\bibnamefont
  {Wenzel}}, \bibinfo {author} {\bibfnamefont {U.}~\bibnamefont {Bandelow}},
  \bibinfo {author} {\bibfnamefont {H.-J.}\ \bibnamefont {W{\"u}nsche}}, \ and\
  \bibinfo {author} {\bibfnamefont {J.}~\bibnamefont {Rehberg}},\ }\href@noop
  {} {\bibfield  {journal} {\bibinfo  {journal} {\textit{IEEE
  J. Quantum Electron.}}\ }\textbf {\bibinfo {volume} {32}},\ \bibinfo {pages} {69}
  (\bibinfo {year} {1996})}\BibitemShut {NoStop}%
\bibitem {Sun2015Dephasing}%
  \BibitemOpen
  \bibfield  {author} {\bibinfo {author} {\bibfnamefont {Y.}~\bibnamefont
  {Sun}}, \bibinfo {author} {\bibfnamefont {Y.}~\bibnamefont {Yang}}, \bibinfo
  {author} {\bibfnamefont {H.}~\bibnamefont {Chen}}, \ and\ \bibinfo {author}
  {\bibfnamefont {S.}~\bibnamefont {Zhu}},\ }\href@noop {} {\bibfield
  {journal} {\bibinfo  {journal} {\textit{Sci. Rep.}}\ }\textbf {\bibinfo
  {volume} {5}} (\bibinfo {year} {2015})}\BibitemShut {NoStop}%
 
\bibitem{Doppler} J. Doppler, A. A. Mailybaev, J. B\"ohm, et al., \textit{Nature} {\bf537}, 76 (2016).
\bibitem{Xu} H. Xu, D. Mason, L. Jiang, and J. G. E. Harris, \textit{Nature} {\bf537}, 80 (2016).
\end{thebibliography}
\end{document}